\documentclass[aps,prl,twocolumn,superscriptaddress,showpacs]{revtex4-1}

\usepackage{amssymb}
\usepackage{amsmath}
\usepackage{graphicx}
\usepackage{dcolumn}
\usepackage{bm}
\usepackage{color}

\begin{document}

\title{Robust Picosecond writing of a Layered Antiferromagnet by Staggered Spin-Orbit-Fields}

\author{P. E. Roy}
\email{per24@cam.ac.uk}
\affiliation{Hitachi Cambridge Laboratory, J J Thomson Avenue,  Cambridge CB3 0HE, UK}

\author{R. M. Otxoa}
\email{ro274@cam.ac.uk}
\affiliation{Hitachi Cambridge Laboratory, J J Thomson Avenue,  Cambridge CB3 0HE, UK}

\author{J. Wunderlich}
\affiliation{Hitachi Cambridge Laboratory, J J Thomson Avenue,  Cambridge CB3 0HE, UK}
\affiliation{Institute of Physics ASCR, v.v.i., Cukrovarnicka 10, 162 53 Praha 6, Czech Republic}


\date{\today}

\begin{abstract}
 Ultrafast electrical switching by current-induced staggered spin-orbit fields, with minimal risk of overshoot, is shown in layered easy-plane antiferromagnets with basal-plane biaxial anisotropy. The reliable switching is due to the field-like torque, relaxing stringent requirements with respect to precision in the time-duration of the excitation pulse. We investigate the switching characteristics as a function of the spin-orbit field strength, pulse duration, pulse rise and fall time and damping by atomistic spin dynamics simulations and an effective equation of motion for the antiferromagnetic order-parameter. The condition, determining the critical spin-orbit field strength for switching is determined and we go on to show that robust picosecond writing is possible at feasible current magnitudes. 
\end{abstract}
\maketitle
The inherent properties of antiferromagnets (AFMs) such as ultrafast dynamics, zero net moment and insensitivity to external magnetic stray fields, make them candidates for a new generation of high speed memory devices. The fact that AFMs also exhibit anisotropic magneto resistance makes it possible to detect the AFM state by electrical means. The writing operation, i.e. reorientation of the AFM sublattice magnetizations has been subject to several proposals, which includes using short laser pulses \cite{Kimel1,Kimel2,Nowak} and spin transfer torque (STT) induced by an impinging spin accumulation generated either by the spin hall effect at a heavy metal/AFM interface \cite{Cheng1} or by injection from a coupled ferromagnet \cite{Gomonay1,Gomonay2}. In those electrical techniques, the spin-accumulation with polarization $\textbf{p}$ causes a staggered field on the AFM sublattices $A$ and $B$  (the staggered nature being what couples effectively to the AFM order parameter) of the form $\textbf{H}_{(A,B)}\sim\textbf{m}_{(A,B)}\times\textbf{p}$. The resulting torque is thus of antidamping type and unless care is taken with respect to pulse duration, an overshoot is a viable risk \cite{Cheng1} as the torque is turned on at all times that the current is on.  Furthermore, several theoretical studies focus on driving AFM domain walls (DWs) by means of STT \cite{Cheng2,Swaving,Braatas1,Braatas2}. These works have been important in predicting the possibility of current-induced excitations in AFMs. However, unless the type of DWs considered in those works can be controllably formed and their position easily detected, it is as of now, difficult to conceive of a device with moving AFM DWs as the mode of operation.  For a robust device, electrical manipulation whereby the AFM order parameter is switched fast and controllably between two stable minima,  without the need for any coupled FMs is a desiriable route to follow. 

In crystals with locally broken inversion-symmetry at the magnetic sites and where  $A$ and $B$ form inversion partners,  another mechanism for AFM spin-axis reorientation presents itself as proposed in ref \cite{Zelenzny}. There, the inverse spin galvanic effect \cite{Ganichev} produces a local non-equilibrium spin polarization, alternating in sign between sublattices $A$ and $B$, thus generating a staggered spin-orbit (SO) field, $\textbf{H}^{\text{SO}}$, which does not depend on $\textbf{m}_{(A,B)}$. $\textbf{H}^{\text{SO}}$ leads then to a field-like torque on $\textbf{m}_{(A,B)}$.  These conditions can be generated by an electrical current density $\textbf{j}$, injected perpendicular to the axis of locally broken inversion symmetry. Proposed  materials to this end are to date $\text{Mn}_{2}\text{Au}$ and CuMnAs\cite{Zelenzny,Sergii,Shick,Barthem,Chun,Jourdan,Maca,Wadley1,Wadley2}. Experimental indication of the electrical manipulation of the AFM state in a multidomain CuMnAs sample has recently been reported \cite{Wadley2}. 
 
\begin{figure}[ht!]
\includegraphics[width=0.5\textwidth]{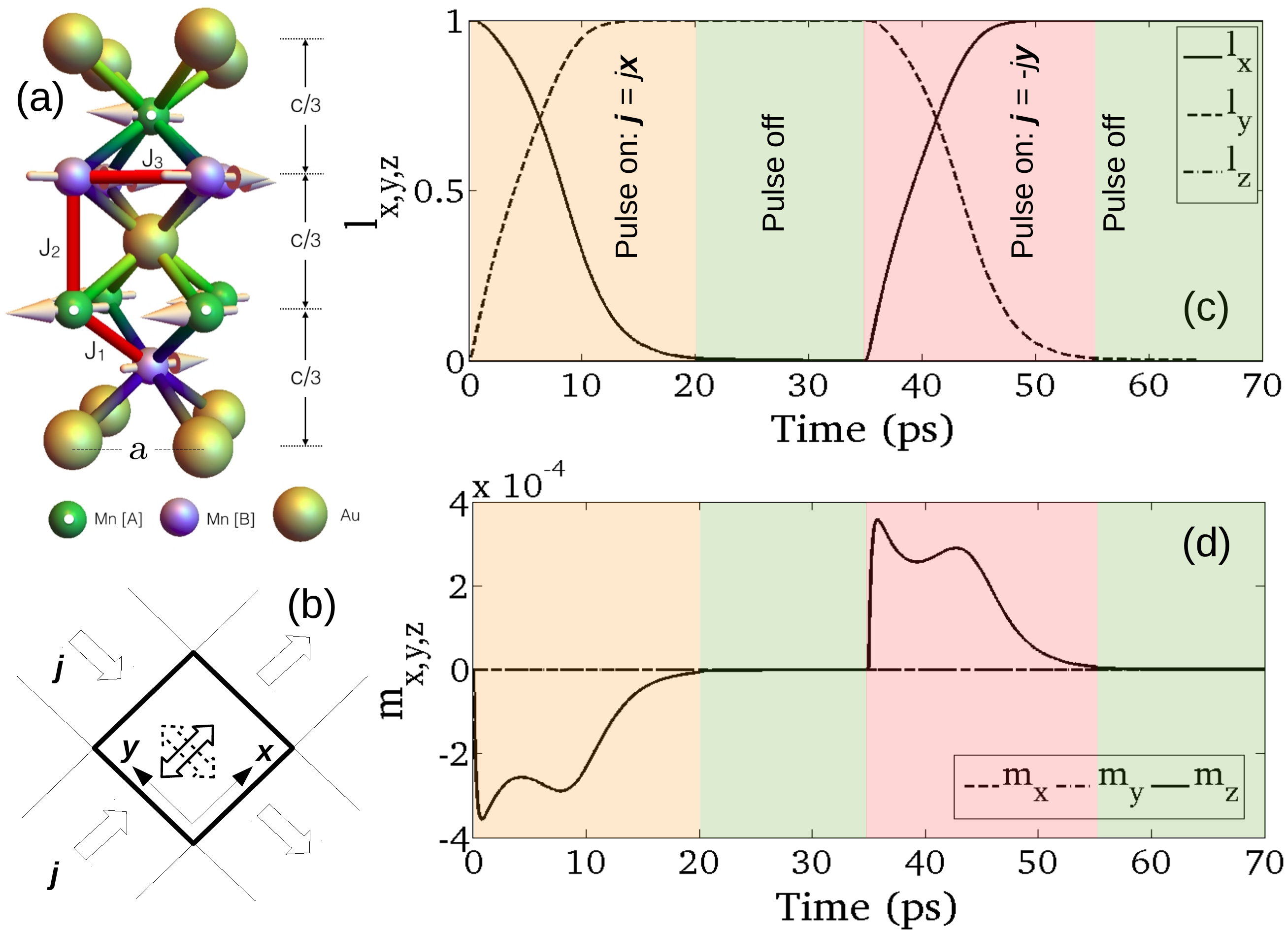}
\caption{(Color online) (a): Crystal and spin structure of $\text{Mn}_{2}\text{Au}$ with basal-plane lattice parameter $a$=3.328 \AA, $c$=8.539 \AA. The bond-exchange constants $J_{1,2,3}$ used are marked by red solid lines. The Mn atoms occupy two types of sites, A and B (see the key).
(b): Coordinate system and orientation of a square device. Current ($\textbf{j}$) injection-directions are indicated by large hollow arrows and two stable positions of the antiferromagnetic sublattices are shown by double arrows in the device. $+\hat{\textbf{z}}$ is along the outward paper normal. (c) and (d): Atomistic spin dynamics results of the time evolution of $\textbf{l}$ (c) and $\textbf{m}$ (d) during two writing operations as described in the text.}
\label{fig:fig1}
\end{figure}

In this work, we show reliable ultrafast switching of a $\text{Mn}_{2}\text{Au}$ device whose body centered tetragonal crystal structure \cite{Wells} is shown Fig.~\ref{fig:fig1}(a). This system exhibits a magnetically hard-axis along the c-axis and has a biaxial anisotropy in the basal-planes with easy directions along the [110] and [1$\overline{1}$0] axes \cite{Zelenzny,Barthem2}. Mn atoms occupy sublattices $A$ and $B$ (Fig.~\ref{fig:fig1}(a)). Typical basal-plane domain sizes in $\text{Mn}_{2}\text{Au}$ is according to ref \cite{Jourdan} $\sim$ 500 nm. Thus for a homogeneous Neel ordered state the lateral dimensions of a thin film device should be smaller than this.  A current injected parallel to the basal planes generates a staggered SO-field, alternating in sign as $\textbf{H}^{\text{SO}}_{A}\sim+\hat{\textbf{z}}\times\textbf{j}$ (at sublattice $A$) and $\textbf{H}^{\text{SO}}_{B}\sim-\hat{\textbf{z}}\times\textbf{j}$ (at sublattice $B$). The resulting torques are therefore field-like. To effectively switch the spin-axes of the sublattices between two stable minima, the biaxial easy directions should coincide with  the current-directions. We thus consider the geometry in  Fig.~\ref{fig:fig1}(b) \cite{Zelenzny}. Recent calculated values of $|\textbf{H}^{\text{SO}}|$ for $\text{Mn}_{2}\text{Au}$ is $\sim\text{20}$ Oe per $\text{10}^{7}\text{A}/\text{cm}^{2}$ (slightly lower than for CuMnAs) \cite{Wadley2}. 

For modeling the device, the total energy, comprised of exchange, tetragonal anisotropy and Zeeman energies is : 
\begin{equation}
\begin{split}
&E=-\sum_{i,j\in N_{i}}J_{ij}\textbf{m}_{i}\cdot \textbf{m}_{j}-K_{2\perp}\sum_{i}\left(\textbf{m}_{i}\cdot\hat{\textbf{u}}_{3}\right)^{2}\\
&\quad-\frac{K_{4\perp}}{2}\sum_{i}\left(\textbf{m}_{i}\cdot\hat{\textbf{u}}_{3}\right)^{4}-\frac{K_{4\parallel}}{2}\sum_{i}\left(\textbf{m}_{i}\cdot\hat{\textbf{u}}_{1}\right)^{4}\\
&\quad-\frac{K_{4\parallel}}{2}\sum_{i}\left(\textbf{m}_{i}\cdot\hat{\textbf{u}}_{2}\right)^{4}-\mu_{0}\mu_{s}\sum_{i}\textbf{m}_{i}\cdot\textbf{H}_{i}^{\text{SO}}
\end{split}
\label{eq:1}
\end{equation}
where $\textbf{m}$ is the unit magnetic moment, $\mu_{0}$ the magnetic permeability in vacuum and $\mu_{s}$  the saturation magnetic moment. The first term in Eq. (1) is the exchange energy with coupling constants  $J_{ij}$ between moments $i$ and $j$. Terms two, three and four constitute the magnetocrystalline anisotropy energy with $K_{2\perp}$,  $K_{4\perp}$ and $K_{4\parallel}$ being the second order perpendicular, fourth order perpendicular and fourth order in-plane anisotropy constants, respectively. Unit vectors $\hat{\textbf{u}}_{1,2,3}$ denote easy directions. Furthermore, $\textbf{H}_{i}^{\text{SO}}$ is the current-induced staggered magnetic SO-field.

\begin{figure}[ht!]
\includegraphics[width=0.5\textwidth]{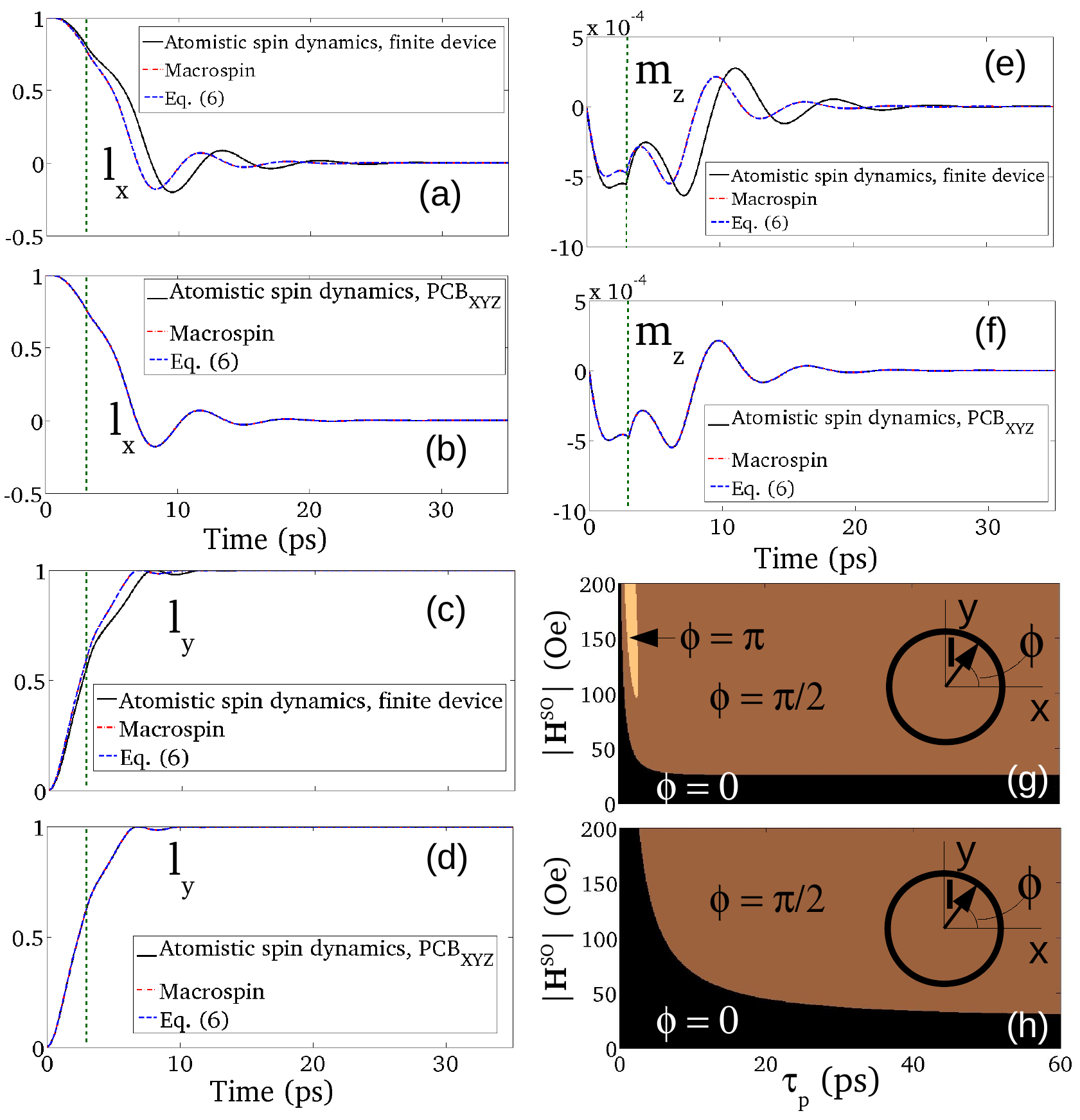}
\caption{ (Color online) (a-f): Correspondence between atomistic spin dynamics, macro-spin modeling and Eq. (6) when comparing to the finite size device used in Fig.1 and when imposing PBCs along $x,y,z$ using 150x150x5 unit cells. (a,b) $l_x$ vs time, (c,d) $l_{y}$ vs time, (e,f) $m_{z}$ vs time. The green vertical dashed lines mark the off-point of the pulse. (g,h): Final angle of $\textbf{l}$ as a function of $|\textbf{H}^{\text{SO}}|$ and $\tau_{p}$ for $\alpha=0.001$ (g) and $\alpha=0.01$ (h). Square pulses have been used in all cases.}
\label{fig:fig2}
\end{figure}

The equation of motion at each site for $\textbf{m}_{i}$ in the precense of the interaction fields $\textbf{H}_{i}$ is given by the Landau-Lifshitz-Gilbert equation:
\begin{equation}
(1+\alpha^{2})\frac{\partial \textbf{m}_{i}}{\partial t}=-\gamma\textbf{m}_{i}\times\textbf{H}_{i}-\alpha\gamma\textbf{m}_{i}\times(\textbf{m}_{i}\times\textbf{H}_{i}).
\label{eq:2}
\end{equation}
Here, $\gamma$ is the gyromagnetic ratio, $\alpha$ is the damping parameter and $\textbf{H}_{i}$ is evaluated from Eq. (\ref{eq:1}) as $\textbf{H}_{i}$ =$\frac{-1}{\mu_{0}\mu_{s}}\frac{\partial E_{i}}{\partial \textbf{m}_{i}}$. We start out by trying the switching capability of the device in  Fig.~\ref{fig:fig1}(b) using atomistic spin dynamics simulations \cite{Evans}. The simulation is carried out for a device of size 150 x 150 x 5 unit cells (49.2 x 49.2 x 4.2695 $\text{nm}^{3}$) of the crystal shown in Fig.~\ref{fig:fig1}(a) . The exchange constants used are $J_{1}$= -396$k_{B}^{-1}$ K, $J_{2}$= -532$k_{B}^{-1}$ K and $J_{1}$= 115$k_{B}^{-1}$ K \cite{Sergii, Barthem, Masrour} where $k_{B}$ is the Boltzmann constant.
For the biaxial basal-plane-ansiotropy, $K_{4\parallel}$  corresponds to an anisotropy field of 100 Oe, as deduced from experiments in ref. \cite{Barthem}; $K_{4\parallel}$ is here then $\text{1.8548}\times\text{10}^{-25}$ J.  $K_{2\perp}$ and  $K_{4\perp}$ per Mn-ion is taken from calculated values in ref \cite{Shick}. We set here $K_{2\perp}$= $\text{-1.303}\times\text{10}^{-22}$ J and use the ratio $K_{4\perp}$=2$K_{4\parallel}$ \cite{Shick}. Here, $\alpha$=0.01 and the Mn magnetic moment, $\mu_{s}$=4$\mu_{b}$ \cite{Barthem}, where $\mu_{b}$ is the Bohr magneton.  Considering the device in  Fig.~\ref{fig:fig1}(b), $\hat{\textbf{u}}_{1}$=$\hat{\textbf{x}}$, $\hat{\textbf{u}}_{2}$=$\hat{\textbf{y}}$ and $\hat{\textbf{u}}_{3}$=$\hat{\textbf{z}}$. Eq. 
 (\ref{eq:2}) is then solved by a fifth order Runge-Kutta scheme \cite{Num}. The first trial consists in applying two current pulses: the first pulse aims to switch the A (B) sublattice from being parallel (antiparallel) to $\hat{\textbf{x}}$ into directions parallel (antiparallel) to $\hat{\textbf{y}}$ and the second pulse to switch the sublattices back to their original state. The procedure is as follows: a  $\tau_{p}$=20 ps long square current pulse is sent along $+\hat{\textbf{x}}$, generating a staggerd $\textbf{H}^{\text{SO}}$ along +$\hat{\textbf{y}}$ on an A-site and along $-\hat{\textbf{y}}$ on a B-site. A waiting time of 15 ps is then imposed to verify the stability of the written state. Then a second current pulse along $-\hat{\textbf{y}}$ is applied, thus generating a staggerd $\textbf{H}^{\text{SO}}$ which is parallel (antiparallel) to $\hat{\textbf{x}}$ on A (B)-sites.  We set $|\textbf{H}^{\text{SO}}|$=100 Oe, corresponding to $\sim$ 5$\times\text{10}^{7}$ A/$\text{cm}^{2}$ . To characterize the state, we use the antiferromagnetic order parameter $\textbf{l}$=$\frac{\textbf{m}_{A}-\textbf{m}_{B}}{2}$ and the magnetization $\textbf{m}$=$\frac{\textbf{m}_{A}+\textbf{m}_{B}}{2}$. As the system is three-dimensional, the volume averaged $\textbf{l}$ and $\textbf{m}$ are extracted. 
As can be seen in Fig.~\ref{fig:fig1}(c), the switching cycle is successful and shows no overshoot in this case. Further, due to the symmetry of the torques, $l_{z}$, $m_{x}$ and $m_{y}$ remain zero at all times. As the SO-torque itself is not staggered, a  build-up in $m_{z}$ occurs (Fig. 1(d)) causing a large precessional exchange torque (first term in  Eq. (\ref{eq:2})) along the $\pm\hat{\textbf{y}}$-directions. Fast switching is therefore the result of an exchange-enhanced torque. The damping exchange-torque (second term in Eq.(\ref{eq:2})) acts as to restore $m_{z}$ to zero, i.e., the lower the $\alpha$, the larger the amplitude of $m_{z}$, which result in shorter switching times and  lower SO-fields required to achieve a switch. Note here that we have used a quite long $\tau_{p}$. In order to achieve a switch, $\tau_{p}$ need only to be long enough to bring $\textbf{l}$ over the biaxial anisotropy barrier, after which even if the pulse is off, the biaxial anisotropy field brings $\textbf{l}$ to the next stable minima. The simulation in  Figs.~\ref{fig:fig1}(c,d) also show that even though $\tau_{p}$ was on for a time longer than that required to cause a switch, $\textbf{l}$ did not overshoot the targeted minmum. The reason is that the the SO-torque is field-like and the direction of $\textbf{H}^{\text{SO}}$ is always along an easy direction for the geometry in Fig. (1b). Thus even for a DC current, $\textbf{l}$ is unlikely to overshoot the targeted anisotropy minima. There may be, however, some conditions, whereby an overshoot event could occur. We therefore firstly investigate the dependence of the final orientation of $\textbf{l}$ for different  $\tau_{p}$ and $|\textbf{H}^{\text{SO}}|$ at different values of $\alpha$ during a switching event. For this, a simplified model allows large sweeps in parameter space at low computational cost:  Thus, we consider a macrospin description where the system consist of two homogeneous antiferromagnetically coupled sublattices carrying total moments $\textbf{m}_{A}$ and $\textbf{m}_{B}$. The current-induced SO-field is then $\textbf{H}^{\text{SO}}$ at $\textbf{m}_{A}$ and $-\textbf{H}^{\text{SO}}$ at $\textbf{m}_{B}$. Setting $J_{ij}<0$, $K_{2\perp}<0$ and using the Gilbert equation, the coupled equations of motion for $\textbf{m}_{A}$ and $\textbf{m}_{B}$ are:
\begin{eqnarray}
\dot{\textbf{m}}_{A} =\omega_{e} [\textbf{m}_{A}\times\textbf{m}_{B}]+\omega_{2\perp}m_{A,z}[\textbf{m}_{A}\times\hat{\textbf{z}}] \nonumber\\
-\omega_{4\perp}m_{A,z}^{3}[\textbf{m}_{A}\times\hat{\textbf{z}}]-\omega_{4\parallel}m_{A,x}^{3}[\textbf{m}_{A}\times\hat{\textbf{x}}]\nonumber\\
-\omega_{4\parallel}m_{A,y}^{3}[\textbf{m}_{A}\times\hat{\textbf{y}}]-\gamma[\textbf{m}_{A}\times\textbf{H}^{\text{SO}}]\nonumber\\
+\alpha\textbf{m}_{A}\times\dot{\textbf{m}}_{A} \label{eq:3}\\
\nonumber\\
\dot{\textbf{m}}_{B} =-\omega_{e} [\textbf{m}_{A}\times\textbf{m}_{B}]+\omega_{2\perp}m_{B,z}[\textbf{m}_{B}\times\hat{\textbf{z}}]  \nonumber\\
-\omega_{4\perp}m_{B,z}^{3}[\textbf{m}_{B}\times\hat{\textbf{z}}]-\omega_{4\parallel}m_{B,x}^{3}[\textbf{m}_{B}\times\hat{\textbf{x}}]\nonumber\\
-\omega_{4\parallel}m_{B,y}^{3}[\textbf{m}_{B}\times\hat{\textbf{y}}]+\gamma[\textbf{m}_{B}\times\textbf{H}^{\text{SO}}]\nonumber\\
+\alpha\textbf{m}_{B}\times\dot{\textbf{m}}_{B}. \label{eq:4}
\end{eqnarray}
Here, $\omega_{e}$=$\frac{\text{2}\gamma |J|}{\mu_{0}\mu_{s}}$ with $|J|=|4J_{1}+J_{2}|$, $\omega_{2\perp}$=$\frac{\text{2}\gamma |K_{2\perp}|}{\mu_{0}\mu_{s}}$, $\omega_{4\perp}$=$\frac{\text{2}\gamma K_{4\perp}}{\mu_{0}\mu_{s}}$ and $\omega_{4\parallel}$=$\frac{\text{2}\gamma K_{4\parallel}}{\mu_{0}\mu_{s}}$ while the dot denotes the time-derivative. $\textbf{l}$ and the total magnetization $\textbf{m}$ are defined as before. It follows that $\textbf{m}\cdot\textbf{l}=0$ and $|\textbf{l}|^{2}+|\textbf{m}|^{2}=1$. Due to strong exchange interaction, we take the exchange limit, $|\textbf{m}|\ll |\textbf{l}|$. Then, $l^{2}\approx 1$ and $\textbf{l}\cdot\dot{\textbf{l}}\approx 0$. The system is describable by $\textbf{m}=(0,0,m_{z})$ and $\textbf{l}=(l_{x},l_{y},0)$ (verified e.g in  Fig.~\ref{fig:fig1}(c,d)). Combining Eqs.(\ref{eq:3})-(\ref{eq:4}) and neglecting the second-order damping terms $\alpha\textbf{m}\times\dot{\textbf{m}}$, $\alpha\textbf{m}\times\dot{\textbf{l}}$ and $\alpha\textbf{l}\times\dot{\textbf{m}}$ in combination with $|\omega_{e}|\gg |\omega_{2\perp}|,\text{ }|\omega_{4\perp}|,\text{ }|\omega_{4\parallel}|,\text{ }\gamma|\textbf{l}\cdot\textbf{H}^{\text{SO}}|$, one arrives at $\textbf{m}\approx -\frac{1}{2\omega_{e}}\textbf{l}\times\dot{\textbf{l}}$ and consequently:
\begin{equation}
\textbf{l}\times\{  \ddot{\textbf{l}}-\omega^{2}_{R}[l_{x}^{3}\hat{\textbf{x}}+l_{y}^{3}\hat{\textbf{y}}]-2\omega_{e}\gamma\textbf{H}^{\text{SO}}+2\omega_{e}\alpha\dot{\textbf{l}}\}\approx 0, \label{eq:5}
\end{equation}
where $\omega_{R}=\sqrt{2\omega_{e}\omega_{4\parallel}}$.
In planar cylindrical coordinates $(l_{x},l_{y})=(\text{cos}\phi,\text{sin}\phi)$ the non-rivial solution of Eq.(\ref{eq:5}) is found by solving :
\begin{equation}
\ddot{\phi}+\frac{\omega_{R}^{2}}{4}\text{sin}(4\phi)-2\omega_{e}\gamma(H^{\text{SO}}_{y}\text{cos}\phi-H^{\text{SO}}_{x}\text{sin}\phi)+2\omega_{e}\alpha\dot{\phi}=0,
\label{eq:6}
\end{equation}
where, if $\textbf{j}\parallel\hat{\textbf{x}}$, then $\textbf{H}^{\text{SO}}$=$H_{y}^{\text{SO}}\hat{\textbf{y}}$ and if $\textbf{j}\parallel-\hat{\textbf{y}}$ then $\textbf{H}^{\text{SO}}$=$H_{x}^{\text{SO}}\hat{\textbf{x}}$. Here, $m_{z}\approx -\frac{1}{2\omega_{e}}\dot{\phi}$. We have in Figs. 2(a-f) included an example of a comparison between the macrospin description (Eqs.(\ref{eq:3})-(\ref{eq:4})), Eq. (\ref{eq:6}) and full atomistic spin dynamics simulations. There,  a low $\alpha=0.001$, $\tau_{p}=3 $ ps and $|\textbf{H}^{\text{SO}}|=40$ Oe (close to the limit of a successful switch) was used as a severe test. Two cases were considered in the atomistic spin dynamics simulations; a finite sized device (same size as that used for the results in  Fig.~\ref{fig:fig1}(c,d)) and periodic boundary conditions (PBC) along $x,y,z$  still with 150 x 150 x 5 unit cells. Firstly, notice that the macro-spin approximation and Eq. (\ref{eq:6}) are in excellent agreement.  Also, there is nearly a perfect overlap of $l_{x,y}$ and $m_{z}$ with the atomistic spin dynamics simulations when PBCs are used (Fig.~\ref{fig:fig2}(b,d,f)). However, quantitative deviations are seen when comparing to the atomistic spin dynamics simulations of a finite sized device (Fig.~\ref{fig:fig2}(a,c,e)), meaning that for this device-size the rotation is not perfectly coherent. In this case the deviations are not severe, so we can safely use Eq.(\ref{eq:6}). Further, we find  that the higher the $\alpha$ the better the correspondence with the atomistic simulations of the finite sized device. 

\begin{figure}[ht!]
\includegraphics[width=0.5\textwidth]{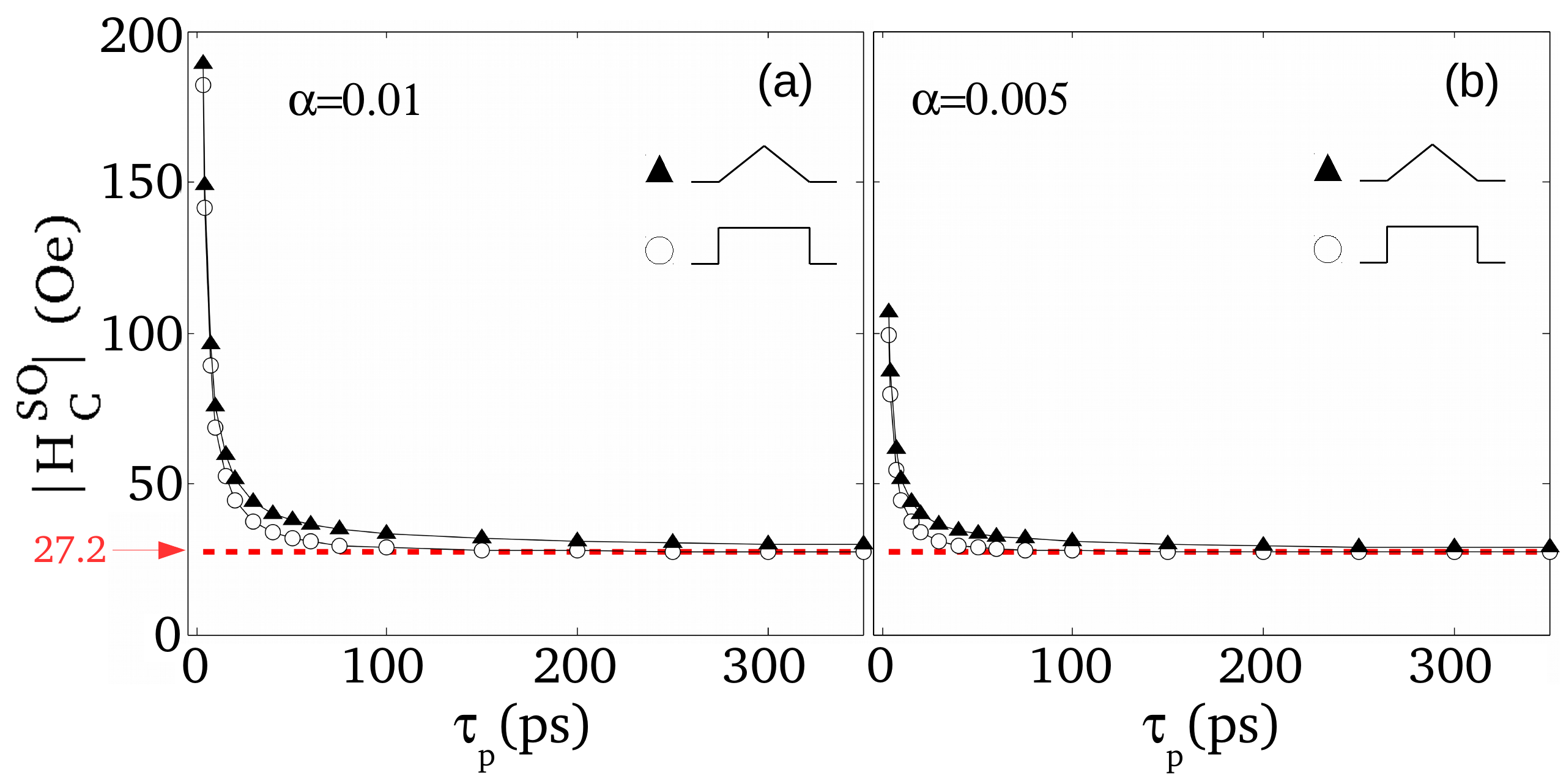}
\caption{ (Color online) (a,b): $|\textbf{H}^{\text{SO}}_{C}|$ vs. $\tau_{p}$ for different $\alpha=0.01$(a) and  $\alpha=0.005$(b). The red dotted line and arrow mark the theoretically lowest $|\textbf{H}^{\text{SO}}_{C}|$. A triangular pulse means that the rise and fall-times equals the pulse duration as defined in  Fig.~\ref{fig:fig4}.}
\label{fig:fig3}
\end{figure}

Relying on Eq. (6) we now investigate the final angle $\phi$ as a function of $|\textbf{H}^{\text{SO}}|$ and $\tau_{p}$ for current pulses $\textbf{j}\parallel\hat{\textbf{x}}$ (i.e. a single switch event). The starting condition is for $\phi=0$ ($\textbf{l}=(1,0)$).
Results are shown in Fig. \ref{fig:fig2}(g,h) for dampings $\alpha=0.001$ and $\alpha=0.01$; a very narrow region of overshoot (Fig.~\ref{fig:fig2}(g)) occurs in the underdamped case ($\alpha=0.001$) for these ranges of $|\textbf{H}^{\text{\text{SO}}}|$ ( $0 <|\textbf{j}|\le\text{10}^{8}\text{A}/\text{cm}^{2}$) and $\tau_{p}$. For $\alpha=0.01$ no overshoot is observed (Fig.~\ref{fig:fig2}(h)).
For a device, a critical parameter is the minimum excitation strength required to write. This, we define as the SO-field required to bring $\textbf{l}$ just over $\phi=\pi/4$, denoted $|\textbf{H}^{\text{SO}}_{C}|$. Applying Eq.(\ref{eq:6}), we calculate $|\textbf{H}^{\text{SO}}_{C}|$ as a function of $\tau_{p}$ for square and triangular pulses considering several $\alpha$. Figs.~\ref{fig:fig3}(a,b) show results for two dampings. 
At short pulse-durations, a $1/\tau_{p}$ dependence can be seen, showing smaller $|\textbf{H}^{\text{SO}}_{C}|$ the lower the $\alpha$ . There is however a minimum  $|\textbf{H}^{\text{SO}}_{C}|$-value which is $\alpha$-independent: invoking $m_{z}\approx -\frac{1}{2\omega_{e}}\dot{\phi}$ and inserting into Eq. (\ref{eq:6}) gives a torque balance equation such that $\frac{\text{d}m_{z}}{\text{dt}}+2\omega_{e}\alpha m_{z}=\frac{\omega_{4}}{4}\text{sin}(\text{4}\phi)-\gamma H_{y}^{\text{SO}}\text{cos}(\phi)$. In the limit of long $\tau_{p}$ with low current amplitude and/or long pulse rise-times, $\text{d}m_{z}/\text{dt}\approx0$ and $2\omega_{e}\alpha m_{z}$ is small compared to the anisotropy and SO-field torques. Thus $\frac{\omega_{4}}{4}\text{sin}(\text{4}\phi)-\gamma H_{y}^{\text{\text{SO}}}\text{cos}(\phi)\approx 0$. As the requirement for a switch is that $\textbf{l}$ just overcomes the anisotropy barrier, it suffices to find the smallest $H_{y}^{\text{SO}}$ on the interval $0\le\phi\le\pi/4$, whereby $\gamma H_{y}^{\text{\text{SO}}}\text{cos}(\phi)\ge\frac{\omega_{4}}{4}\text{sin}(4\phi)$ is satisfied. This yields the lower limit for $H_{y}^{\text{SO}}$=27.2 Oe ($\text{1.36}\times\text{10}^{\text{7}}\text{}\text{A}/\text{cm}^{2}$), which is in excellent agreement to the limits observed in Fig.~\ref{fig:fig3} (horizontal dashed red line).

\begin{figure}[ht!]
\includegraphics[width=0.48\textwidth]{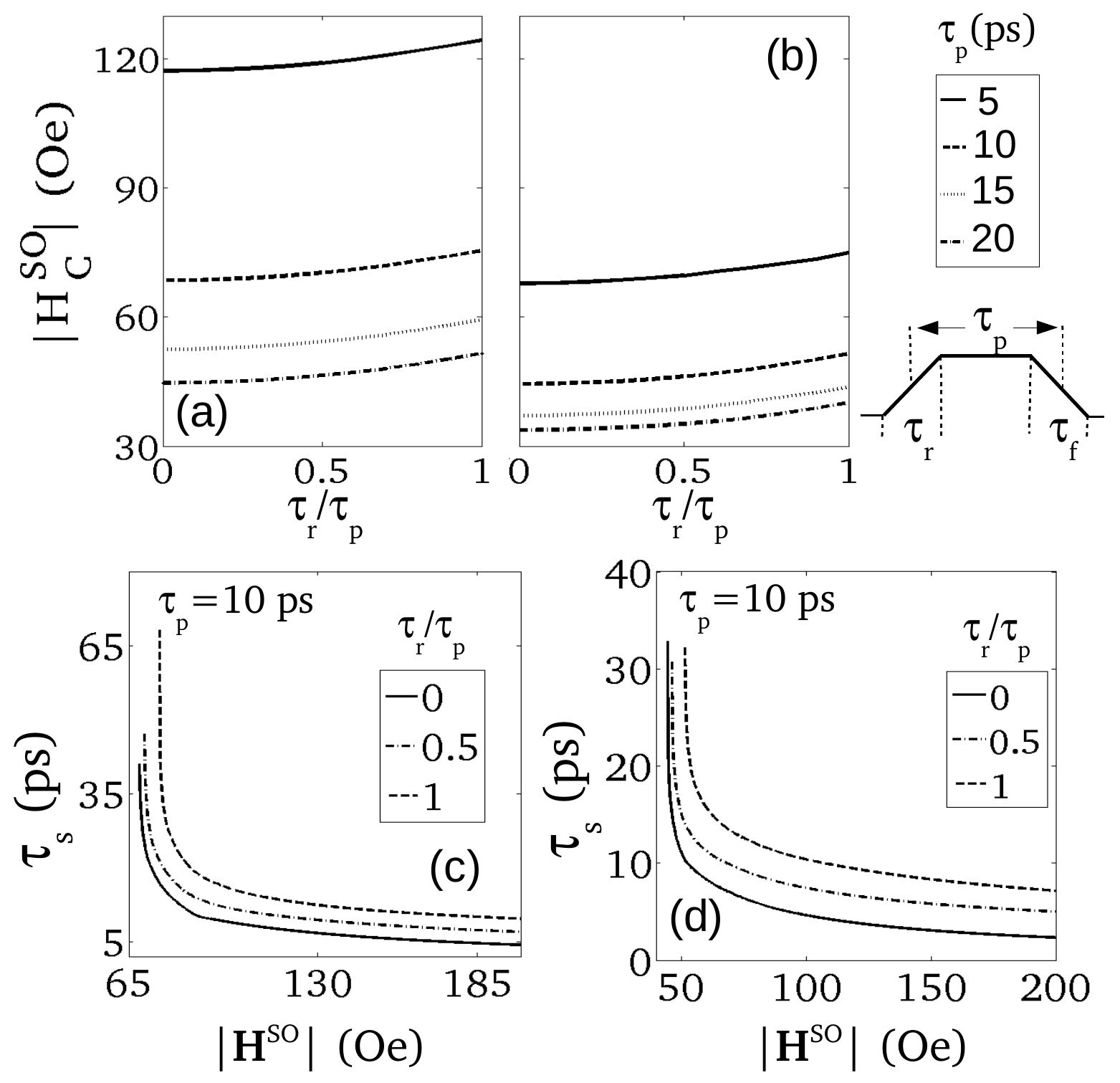}
\caption{(a,b): Dependence of $|\textbf{H}^{\text{SO}}_{C}|$ on $\tau_{r}/\tau_{p}$ for different values of $\tau_{p}$ and  $\alpha$; $\alpha=0.01,0.005$ in (a),(b), respectively. The legend for (a,b) and pulse shape specification is shown to the right ($\tau_{r}/\tau_{p}=0$ means a square pulse and $\tau_{r}/\tau_{p}=1$ is a triangular pulse). (c,d): $\tau_{s}$ versus $|\textbf{H}^{\text{SO}}|$  for different $\tau_{r}/\tau_{p}$ and a fixed $\tau_{p}$=10 ps.
In (c), $\alpha=0.01$ and in (d), $\alpha=0.005$.}
\label{fig:fig4}
\end{figure}

We now investigate the effect of finite pulse rise and fall times, $\tau_{r}$, $\tau_{f}$, respectively by  trapezoidal pulses (as defined in  Fig.~\ref{fig:fig4})  where $\tau_{r}=\tau_{f}$ and $\tau_{p}$.  $|\textbf{H}^{\text{SO}}_{C}|$ is then calculated according to Eq. (\ref{eq:6}) as a function of the ratio $\tau_{r}/\tau_{p}$, considering four different $\tau_{p}$. In Fig.~\ref{fig:fig4}(a) results are shown for $\alpha=0.01$ and in Fig.~\ref{fig:fig4}(b), $\alpha=0.005$.  As can be seen, as far as $|\textbf{H}^{\text{SO}}_{C}|$ is concerned, the dependence on $\tau_{r}/\tau_{p}$ is not severe. Thus, in a real device, pulse-shaping is not crucial to achieve a switching event while keeping injected current magnitudes at a feasible level; e.g. even a $\tau_{r}$=$\tau_{p}$=10 ps (triangular) pulse can switch the device with  $|\textbf{H}^{\text{SO}}_{C}|\sim$ 45-60 Oe, meaning $\sim \text{2.25-3}\times\text{10}^{7}\text{A}/\text{cm}^{2}$ if $\alpha=0.005-0.01$. The reason for a higher $|\textbf{H}^{\text{\text{SO}}}_{C}|$ as $\tau_{r}/\tau_{p}$ increases is a lower maximum amplitude $m_{z}$. The result is a reduced exchange torque.  In terms of the switching time, $\tau_{s}$, defined here as the time it takes for $l_{y}$ to reach $\text{90}\%$ of its maximum value of 1, the difference can be significant. Fig.~\ref{fig:fig4}(c,d) show $\tau_{s}$ versus $|\textbf{H}^{\text{SO}}|$ for three $\tau_{r}/\tau_{p}$-values under a current pulse of $\tau_{p}=\text{10 ps}$. Here two cases are shown in terms of damping; $\alpha=0.01$ and $\alpha=0.005$. As long as one is reasonably above $|\textbf{H}^{\text{SO}}_{C}|$, the behaviour is fairly consistent, with a doubling of the switching time as  $\tau_{r}/\tau_{p}$ approaches 1. The cause is the same as for $|\textbf{H}^{\text{SO}}_{C}|$ although the effect of lower exchange torque is felt throughout the whole switch event ( in determining $|\textbf{H}^{\text{SO}}_{C}|$ only the time between $0\le\phi\le\pi/4$) is relevant. For application point of view, however, the increase in $\tau_{s}$ may not be critical as all switching times are still in the picosecond regime.

In conclusion, we have computationally shown reliable picosecond writing in antiferromagnetic systems whose symmetry allows for current-induced staggered SO-fields. A minimal risk of overshoot due to the field-like torque offers an advantage over structures relying on the antidamping torque. Conditions for the lower limit of the switching field has been found. $|\textbf{H}^{\text{SO}}_{C}|$ has a rather weak dependece on the rise/fall-time of the excitation while $\tau_{s}$ can increase up to a factor of two as the pulse shape goes from rectangular to triangular. The switching times are still in the picosecond time regime. Thus the device remains ultrafast also for non-square pulse shapes.    

P. E. Roy and R. Otxoa contributed equally to this work.

\bibliographystyle{apsrev}

\begin{thebibliography}{1}
\bibitem{Kimel1}
A. V. Kimel, B. A., Ivanov, R. V. Pisarev, A. Kirilyuk and Th. Rasing, Nat. Phys., {\bf 320}, 727 (2009).
\bibitem{Kimel2}
A. V. Kimel, A. Kirilyuk, P. A. Usachev, R. V. Pisarev and Th. Rasing, Nature, {\bf 435}, 655 (2005).
\bibitem{Nowak}
S. Wienholdt, D. Hinzke and U. Nowak, Phys. Rev. Lett., {\bf 108}, 247207 (2012).
\bibitem{Cheng1}
Ran. Cheng, Matthew W. Daniels, Jiang-Gang-Zhu and Di Xiao, Phys. Rev. B., {\bf 91}, 064423 (2015).
\bibitem{Gomonay1}
Helen V. Gomonay and Vadim M. Loktev, Phys. Rev. B., {\bf 81}, 144427 (2010).
\bibitem{Gomonay2}
E. V. Gomonay and V. M. Loktev, Low Temperature Physics, {\bf 40}, 17 (2014).
\bibitem{Cheng2}
Ran Cheng and Qian Niu, Phys. Rev. B., {\bf 89}, 081105(R) (2014).
\bibitem{Swaving}
A. C. Swaving and R. A. Duine, Phys. Rev. B., {\bf 83}, 054428 (2011).
\bibitem{Braatas1}
Erlend G. Tveten, Alireza Qaiumzadeh, O. A. Treiakov and Arne Braatas, Phys. Rev. Lett., {\bf 110}, 127208 (2013).
\bibitem{Braatas2}
Kjetil M. D. Hals, Yaroslav Tserkovnyak and Arne Braatas, Phys. Rev. Lett., {\bf 106}, 107206 (2011).
\bibitem{Zelenzny}
J. Zelenzny, H. Gao, K. Vyborny, J. Zemen, J. Masek, Aurelien Manchon, J. Wunderlich, Jairo Sinova and T. Jungwirth, Phys. Rev. Lett., {\bf 113}, 157201 (2014).
\bibitem{Ganichev}
S. D. Ganichev, Int. J. Mod. Phys. B., {\bf 22}, 1 (2008).
\bibitem{Sergii}
Sergii Khmelevskyi and Peter Mohn, Appl. Phys. Lett., {\bf 93}, 162503 (2008).
\bibitem{Shick}
A. B. Schick, S. Khmelevskyi, O. N. Mryasov, J. Wunderlich and T. Jungwirth, Phys. Rev. B., {\bf 81}, 212409 (2010).
\bibitem{Barthem}
V. M. T. S. Barthem, C. V. Colin, H. Mayaffre, M. -H. Julien and D. Givord, Nat. Commun., {\bf 4}, 2892 (2013).
\bibitem{Chun}
Han-Chun, Zhi-Min Liao, R. G. Sumesh Sofin, Gen Feng, Xin-Mei Ma, Alexander B. Shick, Oleg N. Mryasov and Igor V. Shvets, Adv. Mater., {\bf 24}, 6374 (2012).
\bibitem{Maca}
F. Maca, J. Masek, O. Stelmakhovych, X. Marti, H. Reichlova, K. Uhlirova, P. Beran, P. Wadley, V. Novak and T. Jungwirth, J. Magn. Magn. Mater., {\bf 324}, 1606 (2012).
\bibitem{Wadley1}
P. Wadley {\it et al.}, Nat. Commun., {\bf 4}, 2322 (2013).
\bibitem{Wadley2}
P. Wadley {\it et al.}, Science, {\bf 351}, 587 (2016).
\bibitem{Jourdan}
M. Jourdan, H. Brauning, A. Sapozhnik, H. -J. Elmers, H. Zabel and M. Klaui, J. Phys. D: Appl. Phys., {\bf 48}, 385001 (2015).
\bibitem{Wells}
P. Wells and J. H. Smith, Acta Cryst., A{\bf 26}, 379, 1970.
\bibitem{Barthem2}
Vitoria M. T. S. Barthem, Claire V. Colin, Richard Haettel, Didier Defeu, Dominique Givord, J. Magn. Magn. Mater., {\bf 406}, 289 (2016).
\bibitem{Evans}
R. F. L. Evans, W. J. Fan, P. Chureemart, T. A. Ostler, M. O. A. Ellis and R. W. Chantrell, J. Phys.: Condens. Matter {\bf 26}, 103202 (2014).
\bibitem{Masrour}
R. Masrour, E. K. Hlil, M. Hamedoun, A. Benyoussef, A. Bouahar, H. Lassri, J. Magn. Magn. Mater., {\bf 393}, 600 (2015).
\bibitem{Miltat}
J. Miltat, G. Albequerque, and A. Thiaville, {\it Spin Dynamics in Confined Magnetic Structures I}, Topics in Applied Physics Vol.
83, edited by B. Hillebrands and K. Ounadjela ͑Springer-Verlag, Berlin, 2002, Chapter 1.
\bibitem{Num}
W. H. Press, B. P. Flannery, S. A. Teukolsky, and W. T. Vettering, {\it Numerical Recipes: The Art of Scientific Computing}, Cambridge University Press, Cambridge, England, 1988.
\end{thebibliography}

\end{document}